\documentclass[useAMS,usenatbib]{mn2e}
\usepackage{times}
\usepackage{graphicx}
\usepackage{subfigure}
\usepackage{psfrag}
\usepackage{amsmath}
\usepackage{amssymb}

\def\reff@jnl#1{{\rm#1\/}}
\def\aj{\reff@jnl{AJ}}                 % Astronomical Journal
\def\araa{\reff@jnl{ARA\&A}}           % Annual Review of Astron and Astrophys
\def\apj{\reff@jnl{ApJ}}               % Astrophysical Journal
\def\apjl{\reff@jnl{ApJ}}              % Astrophysical Journal, Letters
\def\apjs{\reff@jnl{ApJS}}             % Astrophysical Journal, Supplement
\def\ao{\reff@jnl{Appl.Optics}}        % Applied Optics
\def\apss{\reff@jnl{Ap\&SS}}           % Astrophysics and Space Science
\def\aap{\reff@jnl{A\&A}}              % Astronomy and Astrophysics
\def\aapr{\reff@jnl{A\&A~Rev.}}        % Astronomy and Astrophysics Reviews
\def\aaps{\reff@jnl{A\&AS}}            % Astronomy and Astrophysics, Supplement
\def\azh{\reff@jnl{AZh}}               % Astronomicheskii Zhurnal
\def\baas{\reff@jnl{BAAS}}             % Bulletin of the AAS
\def\jrasc{\reff@jnl{JRASC}}           % Journal of the RAS of Canada
\def\memras{\reff@jnl{MmRAS}}          % Memoirs of the RAS
\def\mnras{\reff@jnl{MNRAS}}           % Monthly Notices of the RAS
\def\pra{\reff@jnl{Phys.Rev.A}}        % Physical Review A: General Physics
\def\prb{\reff@jnl{Phys.Rev.B}}        % Physical Review B: Solid State
\def\prc{\reff@jnl{Phys.Rev.C}}        % Physical Review C
\def\prd{\reff@jnl{Phys.Rev.D}}        % Physical Review D
\def\prl{\reff@jnl{Phys.Rev.Lett}}     % Physical Review Letters
\def\pasp{\reff@jnl{PASP}}             % Publications of the ASP
\def\pasj{\reff@jnl{PASJ}}             % Publications of the ASJ
\def\qjras{\reff@jnl{QJRAS}}           % Quarterly Journal of the RAS
\def\skytel{\reff@jnl{S\&T}}           % Sky and Telescope
\def\solphys{\reff@jnl{Solar~Phys.}}   % Solar Physics
\def\sovast{\reff@jnl{Soviet~Ast.}}    % Soviet Astronomy
\def\ssr{\reff@jnl{Space~Sci.Rev.}}    % Space Science Reviews
\def\zap{\reff@jnl{ZAp}}               % Zeitschrift fuer Astrophysik
\def\nat{\reff@jnl{Nature}}            % Nature
\title[Detecting extrasolar planets using Bayesian evidence]
      {Detecting extrasolar planets from stellar radial velocities
        using Bayesian evidence} \author[F. Feroz, S.T. Balan and
        M.P. Hobson] {F.~Feroz\thanks{E-mail:
          f.feroz@mrao.cam.ac.uk}, S.~T.~Balan and
        M.~P.~Hobson\\Astrophysics Group, Cavendish
        Laboratory, JJ Thomson Avenue, Cambridge CB3 0HE, UK\\}

\date{Accepted ---. Received ---; in original form \today}
\pagerange{\pageref{firstpage}--\pageref{lastpage}}
\pubyear{2010}

\begin{document}
\label{firstpage}
\maketitle

\begin{abstract}
Stellar radial velocity (RV) measurements have proven to be a very
successful method for detecting extrasolar planets. Analysing RV data
to determine the parameters of the extrasolar planets is a significant
statistical challenge owing to the presence of multiple planets and
various degeneracies between orbital parameters.  Determining the
number of planets favoured by the observed data is an even more
difficult task. Bayesian model selection provides a mathematically
rigorous solution to this problem by calculating marginal posterior
probabilities of models with different number of planets, but the use
of this method in extrasolar planetary searches has been hampered by
the computational cost of the evaluating Bayesian
evidence. Nonetheless, Bayesian model selection has the potential to
improve the interpretation of existing observational data and possibly
detect yet undiscovered planets. We present a new and efficient
Bayesian method for determining the number of extrasolar planets, as
well as for inferring their orbital parameters, without having to
calculate directly the Bayesian evidence for models containing a large
number of planets. Instead, we work iteratively and at each iteration
obtain a conservative lower limit on the odds ratio for the
inclusion of an additional planet into the model. We apply this method
to simulated data-sets containing one and two planets and successfully
recover the correct number of planets and reliable constraints on the
orbital parameters. We also apply our method to RV measurements of HD
37124, 47 Ursae Majoris and HD 10180. For HD 37124, we confirm that
the current data strongly favour a three-planet system. We find strong
evidence for the presence of a fourth planet in 47 Ursae Majoris, but
its orbital period is suspiciously close to one year, casting doubt on
its validity. For HD 10180 we find strong evidence for a six-planet
system.
\end{abstract}

\begin{keywords}

stars: planetary systems -- stars: individual: HD 37124 -- stars:
individual: 47 Ursae Majoris -- stars: individual: HD 10180 --
techniques: radial velocities -- methods: data analysis -- methods:
statistical

\end{keywords}

%%%%%%%%%%%%%%%%%%%%%%%%%%%%%%%%%%%%%%%%%%%%%%%%%%%%%%%%%
\section{Introduction}\label{sec:intro}
%%%%%%%%%%%%%%%%%%%%%%%%%%%%%%%%%%%%%%%%%%%%%%%%%%%%%%%%%

Extrasolar planetary research has been revitalised in the last decade
and so far more than 500 extrasolar planets have been
discovered. Improvements in the accuracy of RV measurements have made
it possible to detect planets with larger orbital periods and smaller
velocity amplitudes. With the flood of new data, more powerful
statistical techniques are being developed and applied to extract as
much information as possible.  Traditionally, planet
parameters and their uncertainties were obtained by searching for
periodicity in the RV data using the Lomb-Scargle periodogram
(\citealt{1976Ap&SS..39..447L, 1982ApJ...263..835S}) to fix the
orbital period and then estimating other parameters by using
minimisation algorithms.

Recent advances in Marko-Chain Monte Carlo (MCMC) techniques (see
e.g. \citealt{MacKay}) have made it possible for Bayesian techniques
to be applied to extrasolar planetary searches (see
e.g. \citealt{2005ApJ...631.1198G, 2005AJ....129.1706F,
  2007ASPC..371..189F, 2009MNRAS.394.1936B}). Bayesian methods have
several advantages over traditional methods, for example when the data
do not cover a complete orbital phase of the planet. Bayesian
inference also provides a rigorous way of performing model selection
which is required to decide the number of planets favoured by the
data.  The main problem in applying such Bayesian model selection
techniques is the computational cost involved in calculating the
Bayesian evidence (see Sec.~\ref{sec:bayesian}). 

\citet{2007ASPC..371..224C} recently reviewed the state of techniques for model selection from a statistical
perspective and \citet{2007ASPC..371..189F} evaluated the performance of a variety of marginal likelihood
estimators in the extrasolar planet context. \citet{2007MNRAS.381.1607G} found good agreement (within 28\%)
between three estimators: (a) parallel tempering, (b) the ratio estimator, and (c) Restricted Monte Carlo (RMC)
for one and two planet models. However,  for a 3 planet model the three estimators diverged significantly with
the RMC yielding the lowest estimate. \citet{2010MNRAS.403..731G} introduced the Nested Restricted Monte Carlo
(NMRC) estimator, an improvement on the RMC estimator. The NRMC estimator is expected to provide a conservative
lower bound on the Bayesian evidence in higher dimensions. These Bayesian model selection techniques have already
resulted in the discovery of previously unknown planets in existing data-sets, e.g. \cite{2009A&A...496L..13T}
discovered a second planet orbiting HD 11506 and \cite{2010MNRAS.403..731G} reported a third planet orbiting 47
Ursae Majoris using Bayesian analysis. Nevertheless, most of the Bayesian model selection techniques employed so
far in extrasolar planetary searches have relied on estimates of the Bayesian evidence, with uncertain accuracy.
Our aim in this paper is to present a new and efficient method for Bayesian model selection to determine the
number of planets favoured by the data, and estimate their parameters, {\em without} having to calculate directly
the Bayesian evidence for models containing a large number of planets.

The outline of this paper is as follows. We give a brief introduction
to Bayesian inference in Sec.~\ref{sec:bayesian} and describe various
Bayesian object detection techniques in
Sec.~\ref{sec:object_detection}. Our model for calculating radial
velocities is described in Sec.~\ref{sec:RV}.  In
Sec.~\ref{sec:bayes_RV} we describe our Bayesian analysis methodology
including the descriptions of likelihood and prior probability
functions. We apply our method to simulated data in 
in Sec.~\ref{sec:mock}, and to real RV data sets on
HD 37124, 47 Ursae Majoris and HD 10180 in Sec.~\ref{sec:real}.
%\ref{sec:HD37124}, \ref{sec:47UMa} and \ref{sec:HD10180}
%respectively. 
Finally our conclusions are presented in
Sec.~\ref{sec:conclusions}.

%%%%%%%%%%%%%%%%%%%%%%%%%%%%%%%%%%%%%%%%%%%%%%%%%%%%%%%%%
\section{Bayesian inference}\label{sec:bayesian}
%%%%%%%%%%%%%%%%%%%%%%%%%%%%%%%%%%%%%%%%%%%%%%%%%%%%%%%%%

Our planet finding methodology is built upon the principles of
Bayesian inference, and so we begin by giving a brief summary of this
framework. Bayesian inference methods provide a consistent approach to
the estimation of a set of parameters $\mathbf{\Theta}$ in a model (or
hypothesis) $H$ for the data $\mathbf{D}$. Bayes' theorem states that
\begin{equation} 
\Pr(\mathbf{\Theta}|\mathbf{D}, H) = \frac{\Pr(\mathbf{D}|\,\mathbf{\Theta},H)\Pr(\mathbf{\Theta}|H)}{\Pr(\mathbf{D}|H)},
\label{eq:bayes}
\end{equation}
where $\Pr(\mathbf{\Theta}|\mathbf{D}, H) \equiv P(\mathbf{\Theta})$ is the posterior probability distribution of
the parameters, $\Pr(\mathbf{D}|\mathbf{\Theta}, H) \equiv \mathcal{L}(\mathbf{\Theta})$ is the likelihood,
$\Pr(\mathbf{\Theta}|H) \equiv \pi(\mathbf{\Theta})$ is the prior, and $\Pr(\mathbf{D}|H) \equiv \mathcal{Z}$ is
the Bayesian evidence.

In parameter estimation, the normalising evidence factor is usually
ignored, since it is independent of the parameters $\mathbf{\Theta}$,
and inferences are obtained by taking samples from the (unnormalised)
posterior using standard MCMC sampling methods, where at equilibrium
the chain contains a set of samples from the parameter space
distributed according to the posterior. This posterior constitutes the
complete Bayesian inference of the parameter values, and can be
marginalised over each parameter to obtain individual parameter
constraints.

In contrast to parameter estimation problems, for model selection the
evidence takes the central role and is simply the factor required to
normalize the posterior over $\mathbf{\Theta}$:
\begin{equation}
\mathcal{Z} = \int{\mathcal{L}(\mathbf{\Theta})\pi(\mathbf{\Theta})}d^D\mathbf{\Theta},
\label{eq:Z}
\end{equation} 
where $D$ is the dimensionality of the parameter space. As the average
of the likelihood over the prior, the evidence is larger for a model
if more of its parameter space is likely and smaller for a model with
large areas in its parameter space having low likelihood values, even
if the likelihood function is very highly peaked. Thus, the evidence
automatically implements Occam's razor: a simpler theory with compact
parameter space will have a larger evidence than a more complicated
one, unless the latter is significantly better at explaining the data.
The question of model selection between two models $H_{0}$ and $H_{1}$
can then be decided by comparing their respective posterior
probabilities given the observed data set $\mathbf{D}$, as follows
\begin{equation}
R = \frac{\Pr(H_{1}|\mathbf{D})}{\Pr(H_{0}|\mathbf{D})}
  = \frac{\Pr(\mathbf{D}|H_{1})\Pr(H_{1})}{\Pr(\mathbf{D}| H_{0})\Pr(H_{0})}
  = \frac{\mathcal{Z}_1}{\mathcal{Z}_0} \frac{\Pr(H_{1})}{\Pr(H_{0})},
\label{eq:R}
\end{equation}
where $\Pr(H_{1})/\Pr(H_{0})$ is the a priori probability ratio for
the two models, which can often be set to unity but occasionally
requires further consideration. The natural logarithm of the ratio of
posterior model probabilities (sometimes termed the posterior odds
ratio) provides a useful guide to what constitutes a significant
difference between two models:
\begin{equation}
\Delta \ln R = \ln \left[ \frac{\Pr(H_{1}|\mathbf{D})}{\Pr(H_{0}|\mathbf{D})}\right]
=\ln \left[ \frac{\mathcal{Z}_1}{\mathcal{Z}_0}\frac{\Pr(H_{1})}{\Pr(H_{0})}\right].
\label{eq:Jeffreys}
\end{equation}
We summarize the convention usually used for model selection in Table~\ref{tab:Jeffreys}.

\begin{table}
\begin{center}
\begin{tabular}{|c|c|c|c|}
\hline
$|\Delta \ln R|$ & Odds & Probability & Remark \\ 
\hline\hline
$<1.0$ & $\lesssim 3:1$ & $<0.750$ & Inconclusive \\
$1.0$ & $\sim 3:1$ & $0.750$ & Weak Evidence \\
$2.5$ & $\sim 12:1$ & $0.923$ & Moderate Evidence \\
$5.0$ & $\sim 150:1$ & $0.993$ & Strong Evidence \\ \hline
\end{tabular}
\end{center}
\caption{The scale we use for the interpretation of model probabilities.}
\label{tab:Jeffreys}
\end{table}

Evaluation of the multidimensional integral in Eq.~\ref{eq:Z} is a
challenging numerical task. Standard techniques like thermodynamic
integration are extremely computationally expensive which makes
evidence evaluation at least an order of magnitude more costly than
parameter estimation. Some fast approximate methods have been used for
evidence evaluation, such as treating the posterior as a multivariate
Gaussian centred at its peak (see e.g. \citealt{Hobson03}), but this
approximation is clearly a poor one for multimodal posteriors (except
perhaps if one performs a separate Gaussian approximation at each
mode). The Savage-Dickey density ratio has also been proposed (see
e.g. \citealt{trotta05}) as an exact, and potentially faster, means of
evaluating evidences, but is restricted to the special case of nested
hypotheses and a separable prior on the model parameters. Various
alternative information criteria for astrophysical model selection are
discussed by \citet{liddle07}, but the evidence remains the preferred
method.

The nested sampling approach, introduced by \citet{skilling04}, is a
Monte Carlo method targeted at the efficient calculation of the
evidence, but also produces posterior inferences as a
by-product. \citet{feroz08} and \citet{multinest} built on this nested
sampling framework and have recently introduced the {\sc MultiNest}
algorithm which is very efficient in sampling from posteriors that may
contain multiple modes and/or large (curving) degeneracies and also
calculates the evidence. This technique has greatly reduces the
computational cost of Bayesian parameter estimation and model
selection and has already been applied to several model selections
problem in astrophysics (see e.g. \citealt{2008arXiv0810.0781F,
  2009MNRAS.398.2049F, 2009CQGra..26u5003F}). We employ this technique
in this paper.

%%%%%%%%%%%%%%%%%%%%%%%%%%%%%%%%%%%%%%%%%%%%%%%%%%%%%%%%%
\section{Bayesian Object Detection}\label{sec:object_detection}
%%%%%%%%%%%%%%%%%%%%%%%%%%%%%%%%%%%%%%%%%%%%%%%%%%%%%%%%%

To detect and characterise an unknown number of objects in a dataset
the Bayesian purist would attempt to infer simultaneously the full set
of parameters $\Theta = \{N_{\rm obj},\Theta_1, \Theta_2, \cdots,
\Theta_{N_{\rm obj}}, \Theta_{\rm n}\}$, where $N_{\rm obj}$ is the
(unknown) number of objects, $\Theta_{\rm i}$ are the parameters
values associated with the $i$th object, and $\Theta_{\rm n}$ is the
set of (nuisance) parameters common to all the objects.  In
particular, this approach allows for the inclusion of an informative
prior (if available) on $N_{\rm obj}$. The crucial complication
inherent in this approach, however, is that the dimensionality of
parameter space is variable and therefore the analysis method should
be able to move between spaces of different dimensionality. Such
techniques are discussed in \cite{Hobson03}. Nevertheless, due to this
additional complexity of variable dimensionality, the techniques are
generally extremely computationally intensive.

An alternative and algorithmically simpler approach for achieving
virtually the same result `by hand' is instead to consider a {\em
  series} of models $H_{N_{\rm obj}}$, each with a {\em fixed} number
of objects, i.e. with $N_{\rm obj}=0,1,2,\ldots$.  One then infers
$N_{\rm obs}$ by identifying the model with the largest marginal
posterior probability $\Pr(H_{N_{\rm obj}}|\mathbf{D})$.  The
probability associated with $N_{\rm obj} = 0$ is often called the
`null evidence' and provides a baseline for comparison of different
models. Indeed, this approach has been adopted previously in exoplanet
studies (see e.g. \citet{2010MNRAS.403..731G}), albeit using only
lower-bound estimates of the Bayesian evidence for each model.
Assuming that there are $n_{\rm p}$ parameters per object and $n_{\rm
  n}$ (nuisance) parameters common to all the objects, for $N_{\rm
  obj}$ objects, there would be $N_{\rm obj}n_{\rm p}+n_{\rm n}$
parameters to be inferred, Thus, the dimensionality of the problem and
consequently the volume of the parameter space increases almost
linearly with $N_{\rm obj}$. Along with this increase in
dimensionality, the complexity of the problem also increases due to
exponential increase in the number of modes as a result of counting
degeneracy, e.g. for $N_{\rm obj} = 2$ and $\Theta = \{\Theta_1,
\Theta_2, \Theta_{\rm n}\}$ where $\Theta_1$ and $\Theta_2$ are the
parameters values associated with first and second objects
respectively and $\Theta_{\rm n}$ is the set of nuisance parameters,
one would get the same value for the likelihood
$\mathcal{L}(\mathbf{\Theta})$ by just rearranging $\Theta$ as
$\{\Theta_2, \Theta_1, \Theta_{\rm n}\}$ and therefore there should at
least be twice as many modes for $N_{\rm obj} = 2$ than for $N_{\rm
  obj} = 1$.  Similarly there are $n!$ more modes for $N_{\rm obj} =
n$ than for $N_{\rm obj} = 1$. This increase in dimensionality and
severe complexity of the posterior makes it very difficult to evaluate
the Bayesian evidence, even approximately.  In exoplanet analyses, we
have found that {\sc MultiNest} is typically capable of evaluating the
evidence accurately for systems with up to 3 planets. If 4 or more
planets are present, {\sc MultiNest} still maps out the posterior
distribution sufficiently well to obtain reliable parameter estimates,
but can begin to produce inaccurate evidence estimates. Thus, even
this approach to Bayesian object detection is of limited applicability
in exoplanet studies.

If the contributions to the data from each object are reasonably well
separated and the correlations between parameters across objects is
minimal, one can use the alternative approach of setting $N_{\rm obj}
= 1$ (see.  e.g. \citealt{Hobson03, feroz08}) and therefore the model
for the data consists of only a single object. This does not, however,
restrict us to detecting only one object in the data. By modelling the
data in such a way, we would expect the posterior distribution to
possess numerous peaks, each corresponding to the location of one of
the objects. Consequently the high dimensionality of the problem is
traded with high multi-modality in this approach, which, depending on
the statistical method employed for exploring the parameter space,
could potentially simplify the problem enormously.  For an application
of this approach in detecting galaxy cluster from weak lensing
data-sets see \cite{2008arXiv0810.0781F}. Unfortunately, for
extrasolar planet detection using RV, this approach cannot be utilized
as the nature of data itself makes the parameters of different planets
in multi-planet system correlated.

We therefore propose here a new general approach to Bayesian object
detection that is applicable to exoplanet studies, even for systems
with a large number of planets. Motivated by the fact that, as
discussed above and in Sec.~\ref{sec:bayesian}, evaluation of the
evidence integral is a far more computationally demanding procedure
than parameter estimation, we consider a method based on the analysis
of residuals remaining after detection and subsequent inclusion
  in the model of $N_{\rm obj}$ objects from the data, as outlined
below. In what follows, we will simply assume that the prior ratio in
Eq.~\ref{eq:Jeffreys} is unity, so that the posterior odds ratio $R$
coincides with the evidence ratio. In principle, however, one could
adopt a more informative prior ratio given a theory of planet
formation that predicted the probability distribution for the number
of planets.

Our approach to Bayesian object detection is as follows. Let us first
denote the observed (fixed) data by $\mathbf{D} = \{d_1, d_2, \cdots,
d_{\rm M}\}$, with the associated uncertainties being $\{\sigma_1,
\sigma_2, \cdots, \sigma_{\rm M}\}$. In the general case that $N_{\rm obj}
= n$, let us define the random variable $\mathbf{D}_n$ as the data
that would be collected if the model $H_n$ were correct, and also the
random variable $\mathbf{R}_n\equiv \mathbf{D}-\mathbf{D}_n$, which
are the data residuals in this case. If we set 
$N_{\rm obj} = n$ and analyse $\mathbf{D}$ to obtain samples from the posterior
distribution of the model parameters $\Theta$, using {\sc
  MultiNest}, then from these samples it is straightforward to obtain
samples from the posterior distribution of the data residuals
$\mathbf{R}_n$. This is given by
\begin{equation}
\Pr(\mathbf{R}_n|\mathbf{D},H_n) = 
\int \Pr(\mathbf{R}_n|\Theta,H_n)\Pr(\Theta|\mathbf{D},H_n)\,d\Theta,
\label{eqn:residdef}
\end{equation}
where 
\begin{equation}
\Pr(\mathbf{R}_n|\Theta,H_n) = 
\prod_{i=1}^M \frac{1}{\sqrt{2\pi\sigma_i^2}}\exp\left\{-\frac{[D_i-R_i-D_{{\rm
        p},i}(\Theta)]^2}{2\sigma_i^2}\right\},
\label{eqn:residdef2}
\end{equation}
and $\mathbf{D}_{\rm p}(\Theta)$ is the (noiseless) predicted data-set
corresponding to the parameter values $\Theta$. It should be noted
that (\ref{eqn:residdef}) and (\ref{eqn:residdef2}) contain no
approximations. In principle, one could then perform a kernel
estimation procedure on the samples obtained to produce a (possibly
analytic) functional form for $\Pr(\mathbf{R}_n|\mathbf{D},H)$. For
simplicity, we assume here that the residuals are independently
Gaussian distributed with a mean $\langle \mathbf{R}_n \rangle =
\{r_1, r_2, \cdots, r_{\rm M}\}$ and standard deviations
$\{\sigma^{\prime}_1, \sigma^{\prime}_2, \cdots, \sigma^{\prime}_{\rm
  M}\}$ obtained from the samples; we find that this 
is a good approximation.

These residual data $\langle\mathbf{R}_n\rangle$, with associated
uncertainities, can then be analysed with $N_{\rm obj} = 0$, giving
the `residual null evidence' $Z_{\rm r,0}$, which is compared with the
evidence value $Z_{\rm r,1}$ obtained by analysing
$\langle\mathbf{R}_n\rangle$ with $N_{\rm obj} = 1$.  We denote the
natural logarithm of the evidence ratio $Z_{\rm r,1}/Z_{\rm r,0}$
between these two models by $\Delta \ln \mathcal{Z_{\rm r}}$. We are
thus comparing the model $H_0$ that the residual data does not contain
an additional planet to the model $H_1$ in which an additional planet
is favoured.

Our overall procedure is therefore as follows. We first set $N_{\rm
  obj}=1$ and analyse the original data set $\mathbf{D}$. If, in the
analysis of the corresponding residuals data, $H_1$ is favoured over
$H_0$, then the original data $\mathbf{D}$ are analysed with $N_{\rm
  obj} = 2$ and the same process is repeated. In this way, $N_{\rm
  obj}$ is increased in the analysis of the original data
$\mathbf{D}$, until $H_0$ is favoured over $H_1$ in the analysis of
the corresponding residual data.  The resulting value for $N_{\rm
  obj}$ gives the number of objects favoured by the data. This
approach thus only requires the Bayesian evidence to be calculated for
$N_{\rm obj} = 1$ model (and the $N_{\rm obj}=0$ model, which is
trivial); this reduces the computational cost of the problem
significantly. Moreover, in principle, this procedure is exact. The
only approximation made here, for the sake of simplicity, is to assume
that $\Pr(\mathbf{R}_n|\mathbf{D},H_n)$ takes the form of an
uncorrelated multivariate Gaussian distribution.

In adopting this approach, our rationale is that, if the
  $n$-planet model is correct, the corresponding data residuals
  $\mathbf{R}_n$ should be consistent with instrumental noise, perhaps
  including an additional stellar jitter contribution (see
  Section~\ref{sec:RV_like}). In this case, the null hypothesis, $H_0$,
    should be preferred over the alternative hypothesis, $H_1$, since
    the latter supposes that some additional signal, not consistent
    with noise, is present in the data residuals. If $H_1$ is
    preferred, we take this as an indication of further planet
    signal(s) present in the data, and therefore re-analysis the {\em
      original} dataset $\mathbf{D}$ using an $(n+1)$-planet model.
    In this way, we circumvent the problem that the inclusion of an
    additional planet to an $n$-planet model will inevitably affect
    the best-fit parameters for the original $n$-planet subset.

%%%%%%%%%%%%%%%%%%%%%%%%%%%%%%%%%%%%%%%%%%%%%%%%%%%%%%%%%
\section{Modelling Radial Velocities}\label{sec:RV}
%%%%%%%%%%%%%%%%%%%%%%%%%%%%%%%%%%%%%%%%%%%%%%%%%%%%%%%%%

It is extremely difficult to observe planets at interstellar distances
directly, since the planets only reflect the light incident on them
from their host star and are consequently many times fainter.
Nonetheless, the gravitational force between the planets and their
host star results in the planets and star revolving around their
common centre of mass. This produces doppler shifts in the spectrum of
the host star according to its RV, the velocity along the
line-of-sight to the observer. Several such measurements, usually over
an extended period of time, can then be used to detect extrasolar
planets.

Following the formalism given in \cite{2009MNRAS.394.1936B}, for
$N_{\rm p}$ planets and ignoring the planet-planet interactions, the
RV at an instant $t_{\rm i}$ observed at $j$th observatory can be
calculated as:
\begin{equation}
v(t_{\rm i},j) = V_{\rm j} - \sum_{\rm p=1}^{N_{\rm p}} K_{\rm p} \left[\sin(f_{\rm i, p} +
  \varpi_{\rm p}) 
+ e_{\rm p} \sin(\varpi_{\rm p})\right],
\label{eq:RV}
\end{equation}
where
\begin{eqnarray*}
V_{\rm j} & = & \mbox{systematic velocity with reference to $j$th observatory},\\
K_{\rm p} & = & \mbox{velocity semi-amplitude of the $p$th planet},\\
\varpi_{\rm p} & = & \mbox{longitude of periastron of 
the $p$th planet},\\
f_{\rm i, p} & = & \mbox{true anomaly of the $p$th planet},\\
e_{\rm p}& = & \mbox{orbital eccentricity of the $p$th planet},\\
P_{\rm p} & = & \mbox{orbital period of the $p$th planet},\\
\chi_{\rm p} & = & \mbox{fraction of an orbit of the $p$th
  planet, prior to the} \\ [-1mm]
& & \mbox{start of data taking, at which
periastron occurred.}
\end{eqnarray*}
Note that $f_{\rm i, p}$ is itself a function of $e_{\rm p}$, $P_{\rm
  p}$ and $\chi_{\rm p}$. While there is unique mean line-of-sight
velocity of the center of motion, it is important to have a different
velocity reference $V_{\rm j}$ for each observatory/spectrograph pair,
since the velocities are measured differentially relative to a
reference frame specific to each observatory.

\begin{table*}
\begin{center}
\begin{tabular}{|c|c|c|c|c|}
\hline
Parameter & Prior & Mathematical Form & Lower Bound & Upper Bound \\ 
\hline\hline
$P$ (days) & Jeffreys & $\frac{1}{P \ln(P_{\rm max}/P_{\rm min})}$ & $0.2$ & $365,000$ \\
$K$ (m/s) & Mod. Jeffreys & $\frac{(K+K_0)^{-1}}{\ln(1+(K_{\rm max}/K_0)(P_{\rm min}/P_{\rm
i})^{1/3}(1/\sqrt{1-e_{\rm i}^2}))}$ & $0$ & $K_{\rm max}(P_{\rm min}/P_{\rm i})^{1/3}(1/\sqrt{1-e_{\rm i}^2})$\\
$V$ (m/s) & Uniform & $\frac{1}{V_{\rm min}-V_{\rm max}}$ & $-K_{\rm max}$ & $K_{\rm max}$ \\
$e$ & Uniform & $1$ & $0$ & $1$ \\
$\varpi$ (rad) & Uniform & $\frac{1}{2\pi}$ & $0$ & $2\pi$ \\
$\chi$ & Uniform & $1$ & $0$ & $1$ \\
$s$ (m/s) & Mod. Jeffreys & $\frac{(s+s_0)^{-1}}{\ln(1+s_{\rm max}/s_0)}$ & $0$ & $K_{\rm max}$ \\ \hline
\end{tabular}
\end{center}
\caption{Prior probability distributions.}
\label{tab:priors}
\end{table*}

We also model the intrinsic stellar variability $s$ (`jitter'), as a
source of uncorrelated Gaussian noise in addition to the measurement
uncertainties. Therefore for each planet we have five free parameters:
$K$, $\varpi$, $e$, $P$ and $\chi$. In addition to these parameters
there are two nuisance parameters $V$ and $s$, common to all the
planets.

These orbital parameters can then be used along with the stellar mass
$m_{\rm s}$ to calculate the length $a$ of the semi-major axis of the
planet's orbit around the centre of mass and the planetary mass $m$ as
follows:
\begin{eqnarray}
a_{\rm s}\sin i & = & \frac{K P \sqrt{1-e^2}}{2 \pi}, \\
\label{eq:as}
m \sin i & \approx &\frac{Km_{\rm s}^{\frac{2}{3}} P^{\frac{1}{3}} \sqrt{1-e^2}}
{(2\pi G)^\frac{1}{3}},
\label{eq:mp}\\
a & \approx & \frac{m_{\rm s} a_{\rm s} \sin i}{m\sin i},
\label{eq:a}
\end{eqnarray}
where $a_{\rm s}$ is the semi-major axis of the stellar orbit about the centre-of-mass and $i$ is the angle
between the direction normal to the planet's orbital plane and the observer's line of sight. Since $i$ cannot be
measured with RV data, only a lower bound on the planetary mass $m$ can be estimated.

%%%%%%%%%%%%%%%%%%%%%%%%%%%%%%%%%%%%%%%%%%%%%%%%%%%%%%%%%
\section{Bayesian Analysis of Radial Velocity Measurements}\label{sec:bayes_RV}
%%%%%%%%%%%%%%%%%%%%%%%%%%%%%%%%%%%%%%%%%%%%%%%%%%%%%%%%%

There are several RV search programmes looking for extrasolar planets. The RV measurements consist of the time
$t_{\rm i}$ of the $i$th observation, the measured RV $v_{\rm i}$ relative to a reference frame and the
corresponding measurement uncertainty $\sigma_{\rm i}$. These RV measurements can be analysed using Bayes'
theorem given in Eq.~\ref{eq:bayes} to obtain the posterior probability distributions of the model parameters
discussed in the previous section. We now describe the form of the likelihood and prior probability
distributions.

%========================================================
\subsection{Likelihood function}\label{sec:RV_like}
%========================================================

As discussed in \cite{2007MNRAS.374.1321G}, the errors on RV
measurements can be treated as Gaussian and therefore the likelihood
function can be written as:
\begin{equation}
\mathcal{L}(\Theta) = \prod_{\rm i} \frac{1}{\sqrt{2\pi(\sigma_{\rm i}^2 + s^2)}} \exp\left[
-\frac{(v(\Theta;t_{\rm i}) - v_{\rm i})^2}{2(\sigma_{\rm i}^2 + s^2)} \right],
\label{eq:L}
\end{equation}
where $v_{\rm i}$ and $\sigma_{\rm i}$ are the $i^{\rm th}$ RV measurement and its corresponding uncertainty
respectively, $v(\Theta;t_{\rm i})$ is the predicted RV for the set of parameters $\Theta$, and $s$ is intrinsic
stellar variability. A large value of $s$ can also indicate the presence of additional planets, e.g. if a
two-planet system is analysed with a single-planet model then the velocity variations introduced by the second
planet would act like an additional noise term and therefore contribute to $s$.

%========================================================
\subsection{Choice of priors}\label{sec:RV_priors}
%========================================================

For parameter estimation, priors become largely irrelevant once the data are sufficiently constraining, but for
model selection the prior dependence always remains. Therefore, it is important that priors are selected based on
physical considerations. We follow the choice of priors given in \cite{2007MNRAS.374.1321G}, as shown in
Table~\ref{tab:priors}.

The modified Jeffreys prior,
\begin{equation}
\Pr(\theta|H) = \frac{1}{(\theta + \theta_0) \ln(1+\theta_{\rm max}/\theta_0)},
\label{eq:modjeff}
\end{equation}
behaves like a uniform prior for $\theta \ll \theta_0$ and like a Jeffreys prior (uniform in $\log$) for $\theta
\gg \theta_0$. We set $K_0 = s_0 = 1$ m/s and $K_{\rm max} = 2129$ m/s, which corresponds to a maximum
planet-star mass ratio of $0.01$.

%%%%%%%%%%%%%%%%%%%%%%%%%%%%%%%%%%%%%%%%%%%%%%%%%%%%%%%%%
\section{Application to Simulated Data}\label{sec:mock}
%%%%%%%%%%%%%%%%%%%%%%%%%%%%%%%%%%%%%%%%%%%%%%%%%%%%%%%%%

In this section, we apply our method to two sets of simulations, one with only one planet in the data and the
other with two planets. Our aim here is to test our new methodology for Bayesian object detection, in particular
the use of the Bayesian evidence in determining the correct number of planets. In particular, we analyse the same
simulations used in \cite{2009MNRAS.394.1936B}, which were obtained by calculating the radial velocities using
(\ref{eq:RV}) for the 1-planet and 2-planet models respectively.  Gaussian noise with $\mu = 0.0$ m/s and $\sigma
= 2.0$ m/s was then added to the resultant radial velocities. 

%========================================================
\subsection{One-planet simulation}\label{sec:sim_analysis}
%========================================================

The evidence and jitter values obtained in the analysis of the 1-planet
simulation are presented in Table~\ref{tab:Z_sim1_Z}. 
\begin{table}
\begin{center}
\begin{tabular}{|c|r|r|r|}
\hline
$N_{\rm p}$ & $\Delta \ln \mathcal{Z}$& $\Delta \ln \mathcal{Z_{\rm r}}$ & $s$ (m/s) \\ 
\hline\hline
$1$ & $82.29 \pm 0.15$ & $-1.33 \pm 0.13$ & $0.42 \pm 0.35$ \\ \hline
\end{tabular}
\end{center}
\caption{The evidence and jitter values for the 1-planet simulation.}
\label{tab:Z_sim1_Z}
\end{table}
Here $\Delta \ln \mathcal{Z}$ denotes the natural logarithm of
evidence ratio $Z_{N_{\rm p}}/Z_0$, where $Z_0$ is the evidence for
$N_{\rm p}=0$. $\Delta \ln \mathcal{Z_{\rm r}}$ is the natural
logarithm of evidence ratio $Z_{N_{\rm r,1}}/Z_{N_{\rm r,0}}$ where
$Z_{N_{\rm r,1}}$ and $Z_{N_{\rm r,0}}$ are the evidence values for
analysing the residual data, after subtracting $N_{\rm p}$ planets, as
discussed in Sec.~\ref{sec:object_detection}, with $1$ and $0$ planets
respectively. $\Delta \ln \mathcal{Z}$ therefore, gives the evidence
in favour of $N_{\rm p}$ planets while $\Delta \ln \mathcal{Z_{\rm
    r}}$ gives the evidence in favour of there being an additional
planet after $N_{\rm p}$ planets have already been found and removed
from the data.  The evidence values listed in Table~\ref{tab:Z_sim1_Z}
should be compared with the scale given in
Table~\ref{tab:Jeffreys}. It is clear that there is overwhelming
evidence for the presence of 1 planet in the data. The negative
$\Delta \ln \mathcal{Z_{\rm r}}$ value further indicates that there is
no evidence for the presence of any additional planets. Furthermore,
the logarithm of the evidence for the 2-planet model was calculated to
be $81.73 \pm 0.16$, which is lower than the logarithm of the evidence
for the 1-planet model listed in Table~\ref{tab:Z_sim1_Z}, providing
further support for the 1-planet model.

Adopting the
1-planet model, therefore, the resulting estimated parameter values
are listed in Table~\ref{tab:Z_sim1_theta} and are in excellent
agreement with the true values used to generate the simulation.

\begin{table}
\begin{center}
\begin{tabular}{|c|r|r|}
\hline
Parameter & True & Estimate \\ 
\hline\hline
$P$ (days) 	& $700.00$	& $705.09 \pm 12.71$	\\
$K$ (m/s) 	& $60.00$	& $60.39 \pm 0.56$	\\
$e$ 		& $0.38$	& $0.38 \pm 0.01$	\\
$\varpi$ (rad) 	& $3.10$	& $3.10 \pm 0.03$	\\
$\chi$	 	& $0.67$	& $0.67 \pm 0.05$	\\
$V$ (m/s) 	& $12.00$	& $11.90 \pm 0.45$	\\
$s$ (m/s) 	& $0.00$	& $0.42 \pm 0.35$	\\ \hline
\end{tabular}
\end{center}
\caption{True and estimated parameter values for the 1-planet simulation.  The estimated values are quoted as
$\mu \pm \sigma$ where $\mu$ and $\sigma$ are the posterior mean and standard deviation respectively.}
\label{tab:Z_sim1_theta}
\end{table}

\subsection{Two-planet simulation}

The evidence and jitter values obtained in the analysis of the 2-planet simulation are presented in
Table~\ref{tab:Z_sim2}. 
\begin{table}
\begin{center}
\begin{tabular}{|c|r|r|r|}
\hline
$N_{\rm p}$ & $\Delta \ln \mathcal{Z}$ & $\Delta \ln \mathcal{Z_{\rm r}}$ & $s$ (m/s) \\ 
\hline\hline
$1$ & $41.92 \pm 0.14$ & $14.82 \pm 0.14$ & $7.47 \pm 1.13$ \\
$2$ & $67.31 \pm 0.16$ & $-1.45 \pm 0.13$ & $0.51 \pm 0.41$ \\ \hline
\end{tabular}
\end{center}
\caption{The evidence and jitter values for the 2-planet simulation.}
\label{tab:Z_sim2}
\end{table}
One can see that for $N_{\rm p} = 1$, the evidence value is quite
large but $\Delta \ln \mathcal{Z_{\rm r}}$ gives a very clear
indication of the presence of an additional planet. The jitter $s$ for
$N_{\rm p} = 1$ is also quite large. The presence of a second planet
is confirmed by $\Delta \ln \mathcal{Z}$ value for $N_{\rm p} = 2$,
which is almost $10$ $\ln$ units higher than for $N_{\rm p} = 1$. The
logarithm of the evidence for the 3-planet model was calculated to be
$66.29 \pm 0.16$, which is lower than the 2-planet model (see
Table.~\ref{tab:Jeffreys}), thus indicating a preference for the
latter. Furthermore, both $\Delta \ln \mathcal{Z_{\rm r}}$ and
$s$ for $N_{\rm p} = 2$ strongly suggest that no additional planet is
present. Thus, adopting the 2-planet model, the estimated parameter
values obtained are listed in Table~\ref{tab:Z_sim2_theta}. Once again
they are in excellent agreement with the true values used to generate
the simulation.

\begin{table}
\begin{center}
\begin{tabular}{|c|r|r|r|r|}
\hline
		& \multicolumn{2}{|c|}{Planet 1}	& \multicolumn{2}{|c|}{Planet 2}	\\
Parameter	& True	& Estimate	& True	& Estimate	\\ 
\hline\hline
$P$ (days) 	& $700.00$	& $708.76 \pm 15.08$	& $100.00$	& $100.45 \pm 0.54$	\\
$K$ (m/s) 	& $60.00$	& $60.35 \pm 0.62$	& $10.00$	& $10.20 \pm 0.62$	\\
$e$ 		& $0.38$	& $0.38 \pm 0.02$	& $0.18$	& $0.19 \pm 0.05$	\\
$\varpi$ (rad) 	& $3.10$	& $3.11 \pm 0.04$	& $1.10$	& $1.27 \pm 0.40$	\\
$\chi$	 	& $0.67$	& $0.67 \pm 0.06$	& $0.17$	& $0.16 \pm 0.08$	\\
$V$ (m/s) 	& $12.00$	& $11.80 \pm 0.52$	&		& $$			\\
$s$ (m/s) 	& $0.00$	& $0.51 \pm 0.41$	&		& $$			\\ \hline
\end{tabular}
\end{center}
\caption{True and estimated parameter values for the 2-planet simulation. The estimated values are quoted as $\mu
  \pm \sigma$ where $\mu$ and $\sigma$ are the posterior mean and standard deviation respectively.}
\label{tab:Z_sim2_theta}
\end{table}

%%%%%%%%%%%%%%%%%%%%%%%%%%%%%%%%%%%%%%%%%%%%%%%%%%%%%%%%%
\section{Application to real data}\label{sec:real}
%%%%%%%%%%%%%%%%%%%%%%%%%%%%%%%%%%%%%%%%%%%%%%%%%%%%%%%%%

In this section, we apply our Bayesian object detection technique to
real RV measurements of HD 37124, 47 Ursae Majoris and HD 10180 and
compare our results with those of previous analyses of these systems.
 
%%%%%%%%%%%%%%%%%%%%%%%%%%%%%%%%%%%%%%%%%%%%%%%%%%%%%%%%%
\subsection{HD 37124}\label{sec:HD37124}
%%%%%%%%%%%%%%%%%%%%%%%%%%%%%%%%%%%%%%%%%%%%%%%%%%%%%%%%%

\begin{figure}
\begin{center}
\includegraphics[width=1\columnwidth]{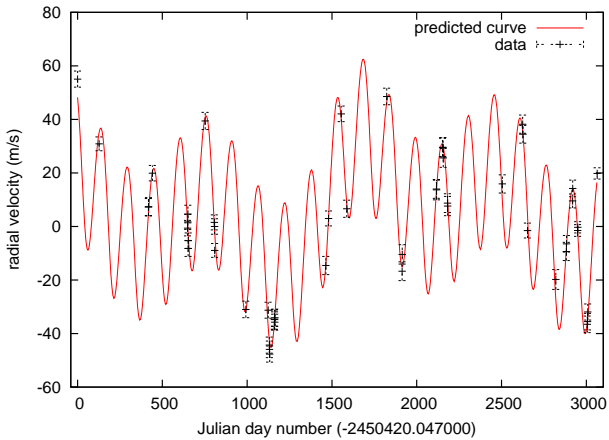}
\caption{Radial velocity measurements, with $1\sigma$ errorbars, and the mean fitted radial velocity curve with
three planets for HD 37124.} 
\label{fig:HD37124_RV}
\end{center}
\end{figure}

HD 37124 is a metal-poor G4 dwarf star at a distance of $33$ pc with
mass $0.85 \pm 0.02$ $M_{\sun}$ \citep{2006ApJ...646..505B,
  2005ApJS..159..141V}. The first planet orbiting HD 37124 was found
by \cite{2000ApJ...536..902V}. Subsequently two further planets were
found by \cite{2003ApJ...582..455B} and \cite{2005ApJ...632..638V}
respectively. We use the 52 RV measurements given in
\cite{2005ApJ...632..638V} for our analysis. The RV data is plotted in
Fig.~\ref{fig:HD37124_RV}.

%%========================================================
%\subsection{Analysis and Results}\label{sec:HD37124_analysis}
%%========================================================

\begin{table}
\begin{center}
\begin{tabular}{|c|r|r|r|}
\hline
$N_{\rm p}$ & $\Delta \ln \mathcal{Z_{\rm r}}$ & $s$ (m/s) \\ 
\hline\hline
$1$ & $12.04 \pm 0.15$ & $13.21 \pm 1.43$ \\
$2$ & $ 5.17 \pm 0.15$ & $ 7.24 \pm 0.93$ \\
$3$ & $-1.62 \pm 0.14$ & $ 2.06 \pm 0.84$ \\ \hline
\end{tabular}
\end{center}
\caption{The evidence and jitter values for the system HD 37124.}
\label{tab:Z_HD37124}
\end{table}

\begin{figure*}
\begin{center}
\includegraphics[width=2\columnwidth]{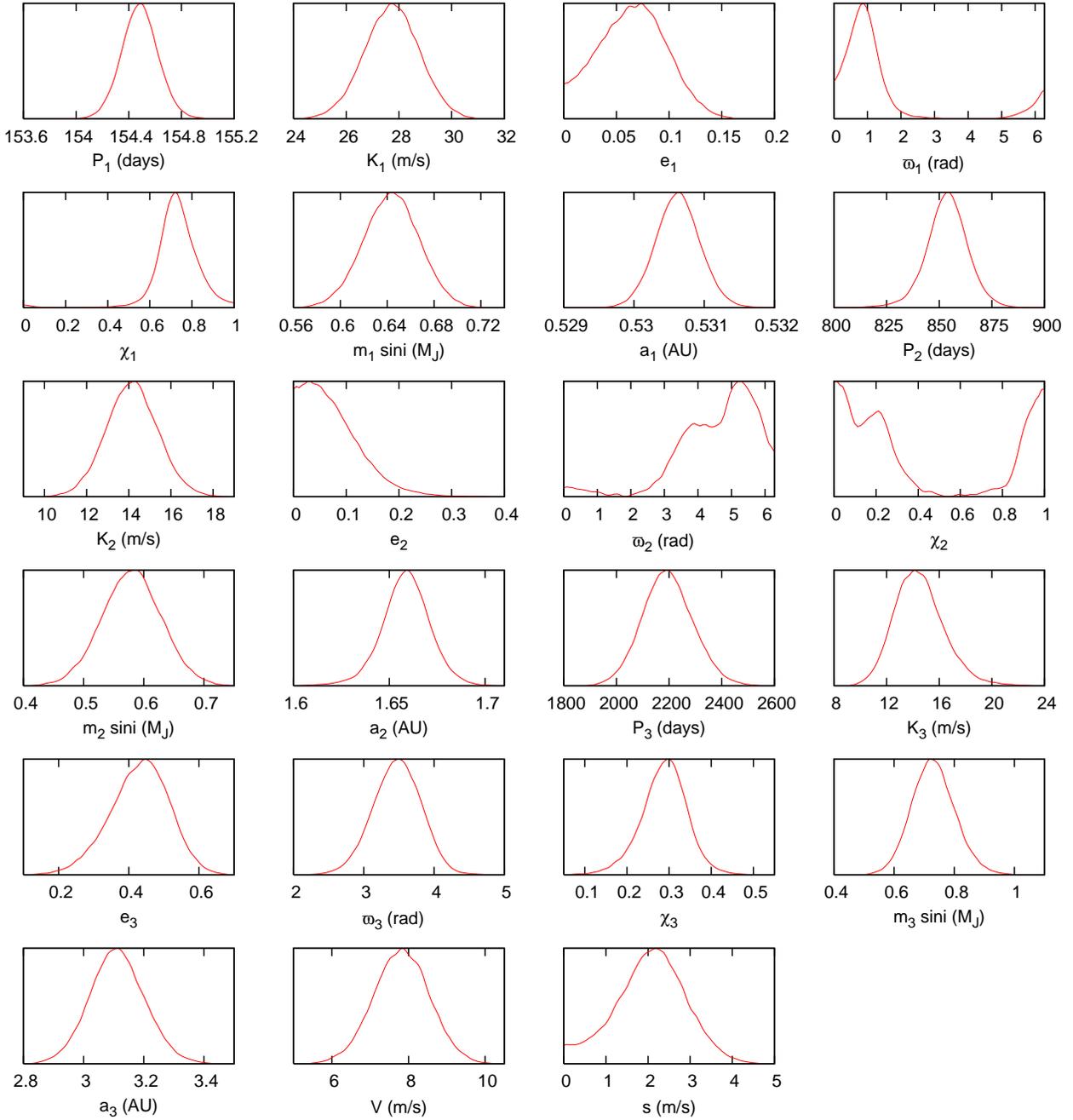}
\caption{1-D marginalised posterior probability distributions for the parameters of the three planets found
orbiting HD 37124.}
\label{fig:HD37124_1D}
\end{center}
\end{figure*}

We follow the object detection methodology outlined in Sec.~\ref{sec:object_detection} and analyse the RV data,
starting with $N_{\rm p} = 1$ and increasing it until the residual evidence ratio $\Delta \ln \mathcal{Z_{\rm r}}
< 0$. The resulting evidence and jitter values are presented in Table~\ref{tab:Z_HD37124}. We can clearly see
$N_{\rm p} = 3$ is the favoured model, with both the residual evidence ratio and jitter values strongly implying
no additional planets are contributing to the data. Adopting the 3-planet model, the estimated parameter values
are listed in Table~\ref{tab:Z_HD37124_theta} while the 1-D marginalised posterior probability distributions are
shown in Fig.~\ref{fig:HD37124_1D}. The mean RV curve for the 3-planet model is overlaid on the RV measurements
in Fig.~\ref{fig:HD37124_RV}.

\begin{table}
\begin{center}
\begin{tabular}{|c|r|r|r|}
\hline
Parameter			& HD 37124 b		& HD 37124 c		& HD 37124 d	      \\
\hline\hline
$P$ (days) 			& $154.48 \pm 0.14$	& $853.70 \pm 10.02$	& $2195.48 \pm 99.06$ \\
	 			&$(154.39)$             & $(855.22)$            & $(2156.73)$         \\
$K$ (m/s) 			& $ 27.73 \pm 1.06$	& $ 14.16 \pm  1.26$	& $  14.52 \pm  1.96$ \\
	 		        &$(28.38)$              & $(14.15)$             & $(14.90)$           \\
$e$ 				& $  0.07 \pm 0.03$	& $  0.08 \pm  0.06$	& $   0.43 \pm  0.09$ \\
	 		        &$(0.10)$               & $(0.04)$              & $(0.45)$            \\
$\varpi$ (rad) 			& $  1.41 \pm 1.57$	& $  4.07 \pm  1.58$	& $   3.47 \pm  0.35$ \\
	 		        &$(0.70)$               & $(5.10)$              & $(3.78)$            \\
$\chi$	 			& $  0.72 \pm 0.13$	& $  0.44 \pm  0.35$	& $   0.29 \pm  0.06$ \\
	 		        &$(0.74)$               & $(0.04)$              & $(0.25)$            \\
$m \sin i$ ($M_{\rm J}$)	& $  0.64 \pm 0.02$	& $  0.58 \pm  0.05$	& $   0.73 \pm  0.07$ \\
	 		        &$(0.66)$               & $(0.58)$              & $(0.75)$            \\
$a$ (AU)			& $  0.53 \pm 0.00$	& $  1.66 \pm  0.01$	& $   3.11 \pm  0.09$ \\
	 		        &$(0.53)$               & $(1.66)$              & $(3.08)$            \\ \hline
\end{tabular}
\end{center}
\caption{Estimated parameter values for the three planets found orbiting HD 37124. The estimated values are
quoted as $\mu \pm \sigma$ where $\mu$ and $\sigma$ are the posterior mean and standard deviation respectively.
The numbers in parenthesis are the maximum-likelihood parameter values.}
\label{tab:Z_HD37124_theta}
\end{table}

Comparing our parameter values with those given in
\cite{2005ApJ...632..638V}, we see that our parameter estimates for
planets HD 37124 b and HD 37124 c are in very good agreement. However,
our orbital time period for HD 37124 d is about 100 days lower and our
estimated eccentricity is somewhat higher. The main reason for this
discrepancy is that \cite{2005ApJ...632..638V} fixed the eccentricity
of HD 37124 d at 0.2 which was chosen to fulfill the dynamical
stability requirement. \cite{2006ApJ...645..688G} also fitted a
3-planet model for HD 37124 and our parameter estimates for all three
planets are in very good agreement with theirs.

%%%%%%%%%%%%%%%%%%%%%%%%%%%%%%%%%%%%%%%%%%%%%%%%%%%%%%%%%
\subsection{47 Ursae Majoris}\label{sec:47UMa}
%%%%%%%%%%%%%%%%%%%%%%%%%%%%%%%%%%%%%%%%%%%%%%%%%%%%%%%%%

\begin{figure}
\begin{center}
\includegraphics[width=1\columnwidth]{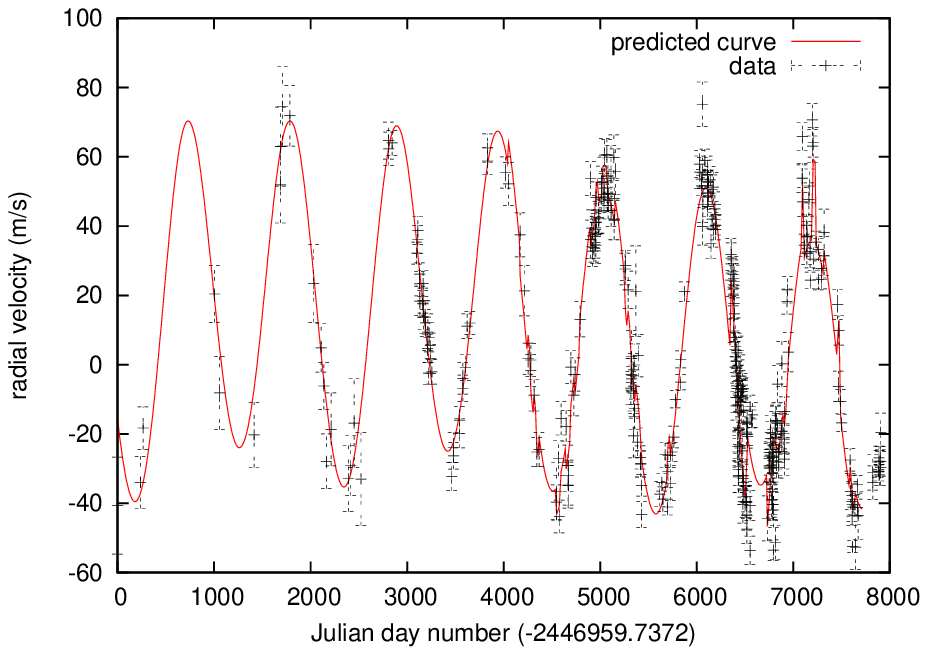}
\caption{Radial velocity measurements, with $1\sigma$ errorbars, and the mean fitted radial velocity curve with three
planets for 47 Ursae Majoris.} 
\label{fig:47UMa_RV}
\end{center}
\end{figure}

47 Ursae Majoris is a solar analog, yellow dwarf star at a distance of
$14.06$ pc with mass $1.06 \pm 0.02$ $M_{\sun}$
\citep{2007ApJS..168..297T}. The first planet orbiting 47 Ursae
Majoris with an orbital period of $1090$ days was found by
\cite{1996ApJ...464L.153B}. A second companion to 47 Ursae Majors with
orbital period of $2594 \pm 90$ days was discovered by
\cite{2002ApJ...564.1028F}. Subsequently the combined RV data for 47
Ursae Majoris from the Lick Observatory, spanning 21.6 years, and from
the 9.2 m Hobbly-Eberly Telescope (HET) and 12.7 m Harlam J.  Smith
(HJS) telescopes of the McDonald Observatory
\citep{2009ApJS..182...97W}, was analysed by
\cite{2010MNRAS.403..731G} and strong evidence was found in favour of
a three-planet system. We analyse the same combined data-set. The RV
data is plotted in Fig.~\ref{fig:47UMa_RV}.

%%========================================================
%\subsection{Analysis and Results}\label{sec:47UMa_analysis}
%%========================================================

\begin{table}
\begin{center}
\begin{tabular}{|c|r|r|}
\hline
$N_{\rm p}$ & $\Delta \ln \mathcal{Z_{\rm r}}$ & $s$ (m/s) \\ 
\hline\hline
$1$ & $98.27 \pm 0.25$ & $10.13 \pm 0.47$ \\
$2$ & $23.32 \pm 0.25$ & $ 6.19 \pm 0.36$ \\
$3$ & $ 4.39 \pm 0.25$ & $ 4.87 \pm 0.33$ \\
$4$ & $-0.77 \pm 0.23$ & $ 4.35 \pm 0.33$ \\ \hline
\end{tabular}
\end{center}
\caption{The evidence and jitter values for the system 47 Ursae Majoris.}
\label{tab:Z_47UMa}
\end{table}

\begin{table*}
\begin{center}
\begin{tabular}{|c|r|r|r|}
\hline
Parameter		& 	47 UMa b	     & 47 UMa c	      	    & 47 UMa d	              \\
\hline\hline
$P$ (days) 		&	$1078.26 \pm 1.83$   & $2293.17 \pm 79.39$  & $14674.55 \pm  5925.37$ \\
	 		        &$(1078.69)$         & $(2228.61)$          & $(17217.04)$            \\
$K$ (m/s) 		&	$  49.49 \pm 1.53$   & $   8.49 \pm  1.30$  & $   13.52 \pm	1.09$ \\
	 		        &$(51.22)$           & $(10.18)$            & $(13.42)$               \\
$e$ 			&	$   0.03 \pm 0.01$   & $   0.32 \pm  0.18$  & $    0.24 \pm	0.16$ \\
	 		        &$(0.04)$            & $(0.55)$             & $(0.36)$                \\
$\varpi$ (rad) 		&	$   4.32 \pm 0.74$   & $   2.95 \pm  1.32$  & $    2.37 \pm	2.37$ \\
	 		        &$(4.29)$            & $(2.42)$             & $(0.32)$                \\
$\chi$	 		&	$   0.39 \pm 0.11$   & $   0.64 \pm  0.28$  & $    0.58 \pm	0.19$ \\
	 		        &$(0.41)$            & $(0.75)$             & $(0.69)$                \\
$m \sin i$ ($M_{\rm J}$)&	$   2.59 \pm 0.09$   & $   0.53 \pm  0.05$  & $    1.58 \pm	0.17$ \\
	 		        &$(2.71)$            & $(0.57)$             & $(1.66)$                \\
$a$ (AU)		&	$   2.10 \pm 0.02$   & $   3.48 \pm  0.08$  & $   11.81 \pm	2.99$ \\
	 		        &$(2.11)$            & $(3.43)$             & $(13.40)$               \\ \hline
\end{tabular}
\end{center}
\caption{Estimated parameter values for the three planets found
  orbiting 47 Ursae Majoris. The estimated values are quoted as $\mu
  \pm \sigma$ where $\mu$ and $\sigma$ are the posterior mean and
  standard deviation respectively. The numbers in parentheses are the
  maximum-likelihood parameter values.}
\label{tab:Z_47UMa_theta}
\end{table*}

\cite{2010MNRAS.403..731G} analysed the RV data this system by
ignoring the residual velocity offsets associated with dewar changes,
as well as by incorporating the dewar velocity offsets as additional
unknown parameters, and found the results to be consistent. We
therefore ignore the velocity offsets associated with dewar changes
and fit for three velocity offsets $V_{\rm L}$, $V_{\rm HET}$ and
$V_{\rm HJS}$ associated with Lick, HET and HJS telescopes
respectively.

We follow the object detection methodology outlined in
Sec.~\ref{sec:object_detection} and analyse the RV data, starting with
$N_{\rm p} = 1$ and increasing it until the residual evidence ratio
$\Delta \ln \mathcal{Z_{\rm r}} < 0$. The resulting evidence and
jitter values are presented in Table~\ref{tab:Z_47UMa}. We can clearly
see $N_{\rm p} = 4$ is the favoured model, with the residual evidence
ratio strongly implying no additional planets are contributing to the
data. Our detection of the fourth planet contradicts the analysis of
\cite{2010MNRAS.403..731G}, which did not find a well-defined peak for
the fourth period using combined Lick, HET and HJS data-sets. They
did, however, find the fourth planet using only the Lick data-set, but
their calculated upper limit on the false alarm probability for the
presence of the fourth planet of $\approx 0.5$ was deemed too
high. Our detected fourth planet has the best-fit orbital period of
$369.7$ days, consistent with the period of fourth planet found by
\cite{2010MNRAS.403..731G} in Lick-only data. Nonetheless, this period
is suspiciously close to one year, indicating that it might be an
artefact of the data reduction. We therefore discuss the results
obtained from the 3-planet model in the rest of this section.

\begin{figure*}
\begin{center}
\includegraphics[width=2\columnwidth]{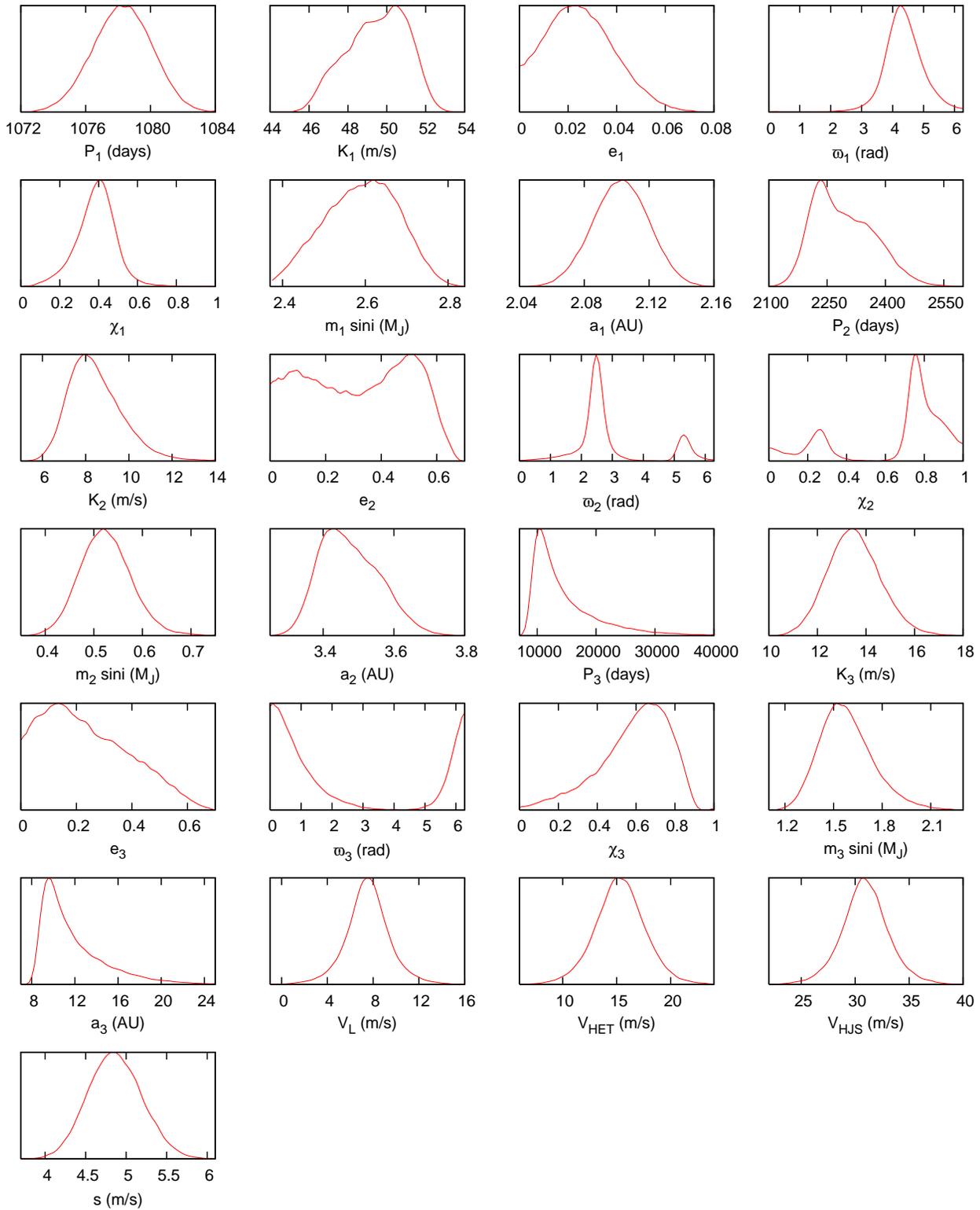}
\caption{1-D marginalised posterior probability distributions for the parameters of the three planets found
orbiting 47 Ursae Majoris.}
\label{fig:47UMa_1D}
\end{center}
\end{figure*}

Adopting the 3-planet model, the estimated parameter values are listed
in Table~\ref{tab:Z_47UMa_theta} while the 1-D marginalised posterior
probability distributions are shown in Fig.~\ref{fig:47UMa_1D}. The
mean RV curve for the 4-planet model is overlaid on the RV
measurements in Fig.~\ref{fig:47UMa_RV}. There is fairly good
agreement between our parameter constraints and those presented by
\cite{2010MNRAS.403..731G}.

%%%%%%%%%%%%%%%%%%%%%%%%%%%%%%%%%%%%%%%%%%%%%%%%%%%%%%%%%
\subsection{HD 10180}\label{sec:HD10180}
%%%%%%%%%%%%%%%%%%%%%%%%%%%%%%%%%%%%%%%%%%%%%%%%%%%%%%%%%

\begin{figure}
\begin{center}
\includegraphics[width=1\columnwidth]{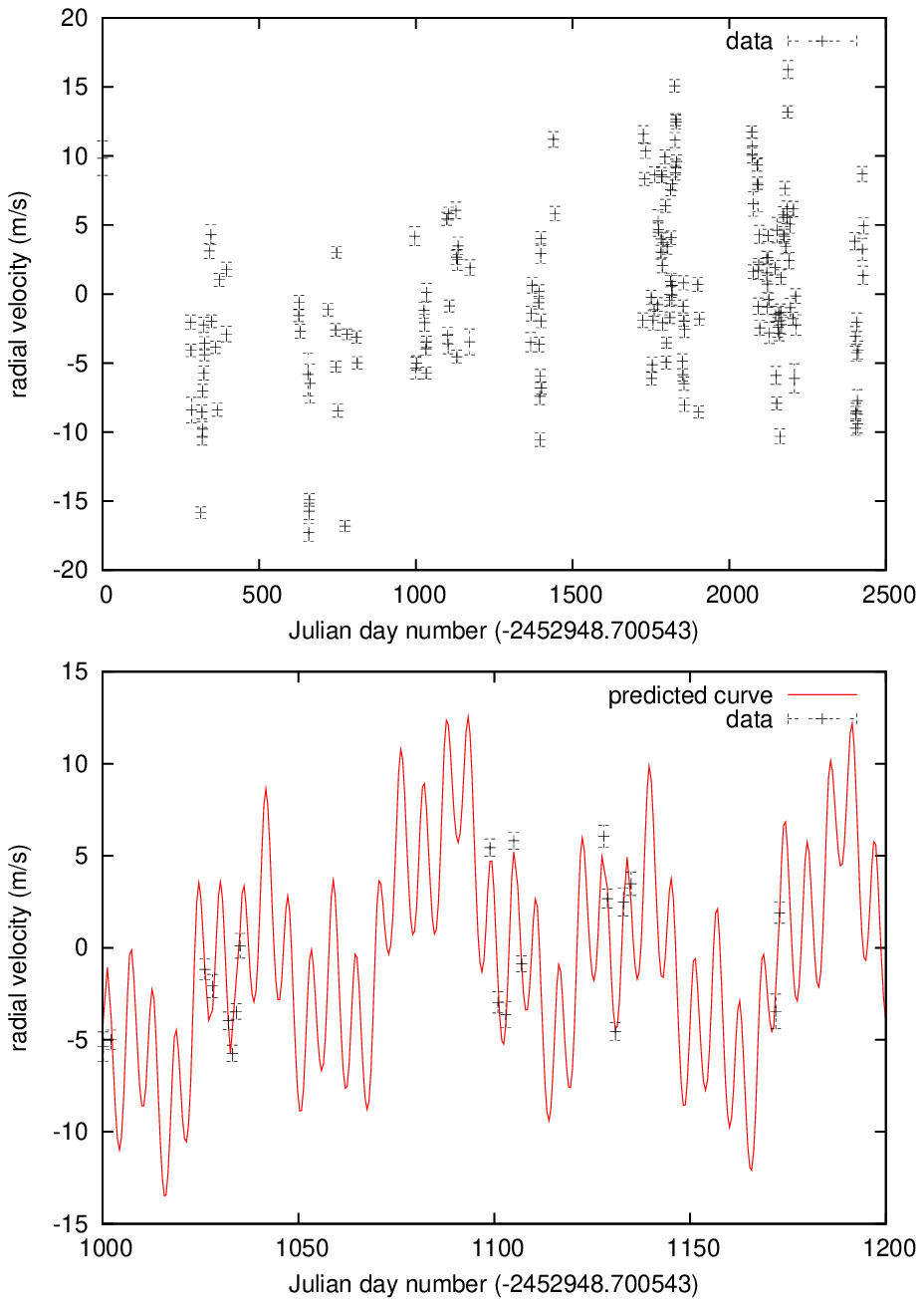}
\caption{Top panel shows the radial velocity measurements (after subtracting mean RV of 3.55302 km/s), with
$1\sigma$ errorbars. Bottom panel shows a blow-up of the mean fitted radial velocity curve with six planets for
HD 10180.}
\label{fig:HD10180_RV}
\end{center}
\end{figure}

HD 10180 is a G1 V type star at a distance of $39$ pc with mass $1.06
\pm 0.05$ $M_{\sun}$ \citep{2010arXiv1011.4994L}. Using the RV data
from HARPS instrument \citep{2003Msngr.114...20M},
\cite{2010arXiv1011.4994L} recently reported at least five and as many
as seven planets orbiting this star. There has been much interest in
the possible seventh planet as its minimum mass as reported by
\cite{2010arXiv1011.4994L} is 1.4 $M_{\earth}$. We analyse the same
HARPS data-set
%\footnote{retrieved electronically from \newline
%  \texttt{http://cdsarc.u-strasbg.fr/cgi-bin/qcat?J/A+A/526/A141}}
after subtracting a mean radial velocity of $3.55302$ km/s from
it. The resultant RV data is plotted in Fig.~\ref{fig:HD10180_RV}.

%%========================================================
%\subsection{Analysis and Results}\label{sec:HD10180_analysis}
%%========================================================

\begin{table}
\begin{center}
\begin{tabular}{|c|r|r|}
\hline
$N_{\rm p}$ & $\Delta \ln \mathcal{Z_{\rm r}}$ & $s$ (m/s) \\ 
\hline\hline
$1$ & $24.84 \pm 0.17$ & $5.64 \pm 0.29$ \\
$2$ & $ 9.46 \pm 0.18$ & $4.55 \pm 0.23$ \\
$3$ & $63.47 \pm 0.17$ & $3.96 \pm 0.20$ \\
$4$ & $45.47 \pm 0.17$ & $2.45 \pm 0.13$ \\ 
$5$ & $ 4.49 \pm 0.17$ & $1.58 \pm 0.09$ \\
$6$ & $-0.73 \pm 0.17$ & $1.36 \pm 0.07$ \\\hline
\end{tabular}
\end{center}
\caption{The evidence and jitter values for the system HD 10180.}
\label{tab:Z_HD10180}
\end{table}

\begin{table*}
\begin{center}
\begin{tabular}{|c|r|r|r|r|r|r|}
\hline
Parameter		& HD 10180 b & HD 10180 c & HD 10180 d & HD 10180 e & HD 10180 f & HD 10180 g \\
\hline\hline
$P$ (days) 		&	$5.76 \pm 0.02$ & $16.35 \pm 0.05$ & $49.74 \pm 0.20$ & $122.75 \pm 0.54$ & $600.17 \pm 13.75$ & $2266.22 \pm 412.42$ \\
	 		        &$(5.76)$ & $(16.36)$ & $(49.74)$ & $(122.69)$ & $(601.88)$ & $(2231.44)$ \\
$K$ (m/s) 		&	$4.54 \pm 0.12$ & $2.89 \pm 0.13$ & $4.28 \pm 0.14$ & $2.91 \pm 0.14$ & $1.43 \pm 0.20$ & $3.06 \pm 0.16$ \\
	 		        &$(4.63)$ & $(2.94)$ & $(4.25)$ & $(2.70)$ & $(1.79)$ & $(2.98)$ \\
$e$ 			&	$0.07 \pm 0.03$ & $0.13 \pm 0.04$ & $0.03 \pm 0.02$ & $0.09 \pm 0.04$ & $0.15 \pm 0.09$ & $0.09 \pm 0.05$ \\
	 		        &$(0.08)$ & $(0.12)$ & $(0.03)$ & $(0.08)$ & $(0.25)$ & $(0.05)$ \\
$\varpi$ (rad) 		&	$2.60 \pm 0.38$ & $2.62 \pm 0.35$ & $2.56 \pm 0.16$ & $2.65 \pm 0.53$ & $3.08 \pm 0.97$ & $2.89 \pm 2.60$ \\
	 		        &$(2.51)$ & $(2.49)$ & $(5.12)$ & $(2.95)$ & $(2.43)$ & $(5.98)$ \\
$\chi$	 		&	$0.22 \pm 0.06$ & $0.35 \pm 0.06$ & $0.43 \pm 0.27$ & $0.23 \pm 0.11$ & $0.31 \pm 0.28$ & $0.67 \pm 0.10$ \\
	 		        &$(0.24)$ & $(0.37)$ & $(0.83)$ & $(0.16)$ & $(0.27)$ & $(0.73)$ \\
$m \sin i$ ($M_{\rm J}$)&	$0.04 \pm 0.00$ & $0.04 \pm 0.00$ & $0.08 \pm 0.00$ & $0.07 \pm 0.00$ & $0.06 \pm 0.00$ & $0.20 \pm 0.01$ \\
	 		        &$(0.04)$ & $(0.04)$ & $(0.08)$ & $(0.07)$ & $(0.07)$ & $(0.20)$ \\
$a$ (AU)		&	$0.06 \pm 0.00$ & $0.13 \pm 0.00$ & $0.27 \pm 0.00$ & $0.49 \pm 0.00$ & $1.42 \pm 0.03$ & $3.45 \pm 0.16$ \\
	 		        &$(0.06)$ & $(0.13)$ & $(0.27)$ & $(0.49)$ & $(1.42)$ & $(3.40)$ \\ \hline
\end{tabular}
\end{center}
\caption{Estimated parameter values for the six planets found orbiting HD 10180. The estimated values are
quoted as $\mu \pm \sigma$ where $\mu$ and $\sigma$ are the posterior
mean and standard deviation respectively. The numbers in parenthesis are
the maximum-likelihood parameter values.}
\label{tab:Z_HD10180_theta}
\end{table*}

\begin{figure*}
\begin{center}
\includegraphics[width=2\columnwidth]{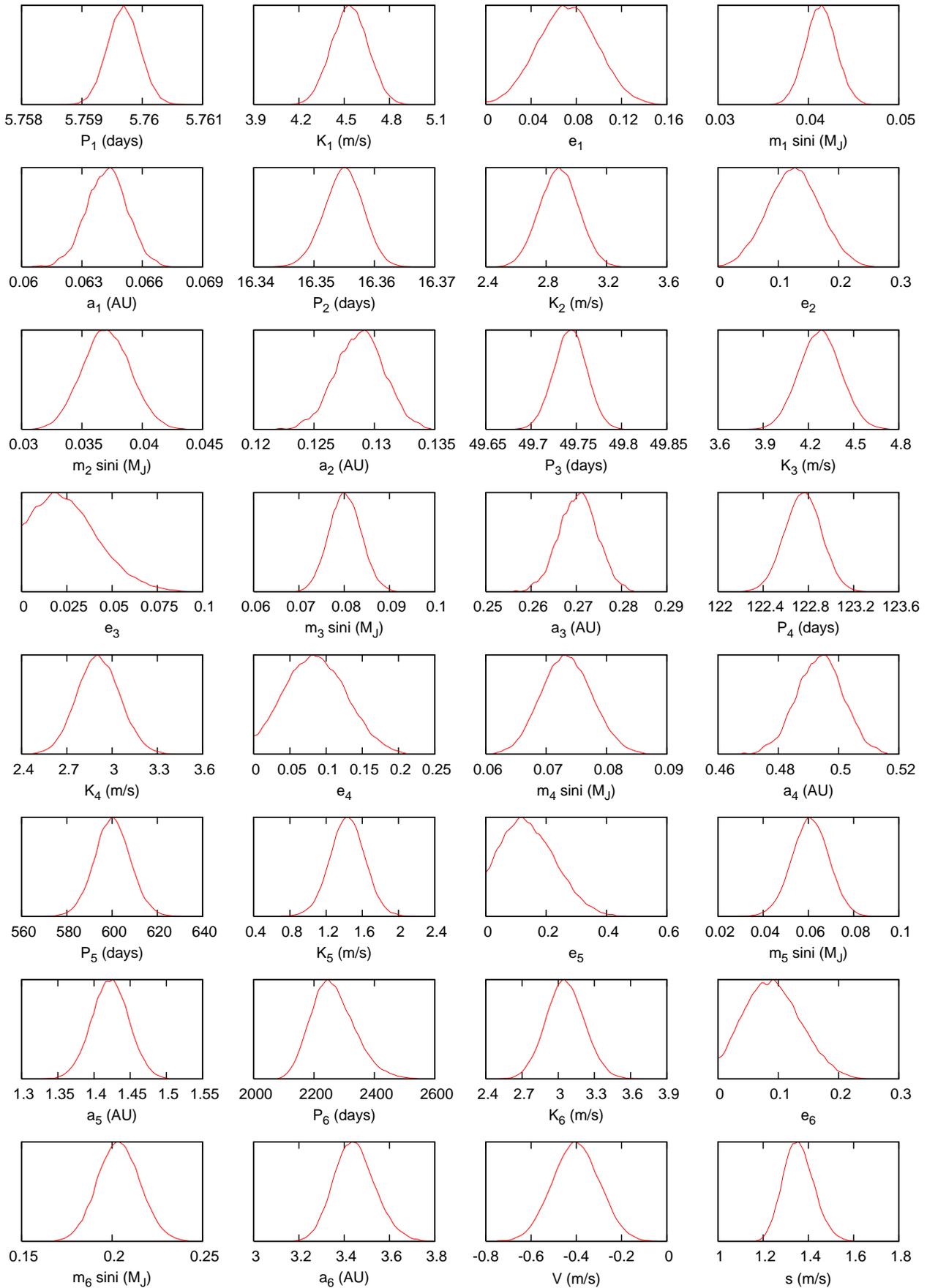}
\caption{1-D marginalised posterior probability distributions for the
  parameters of the six planets found
orbiting HD 10180.}
\label{fig:HD10180_1D}
\end{center}
\end{figure*}

The evidence and jitter values are presented in Table~\ref{tab:Z_HD10180}. We can clearly see $N_{\rm p} = 6$ is
the favoured model, with the residual evidence ratio strongly implying that the residual data consists of noise
only. Adopting the 6-planet model, the estimated parameter values are listed in Table~\ref{tab:Z_HD10180_theta}
while the 1-D marginalised posterior probability distributions are shown in Fig.~\ref{fig:HD10180_1D}. The mean
RV curve for the 6-planet model is overlaid on the RV measurements in Fig.~\ref{fig:HD10180_RV}. It can be seen
that our orbital parameters are in general reasonably good agreement with the ones presented in
\cite{2010arXiv1011.4994L}.

\cite{2010arXiv1011.4994L} found fairly strong peaks with periods
1.178 and 6.51 days in the periodogram of the residuals of the
6-planet Kaplerian model. They noted that these two peaks are aliases
of each other with 1 sidereal day period $(|1/6.51 - 1.0027| \approx
1/1.178)$. Arguing that it is unlikely for the system to be
dynamically stable with two planets having $P = 5.76$ days and $P =
6.51$ days, they concluded that if the 7th signal is caused by a
planet, it is likely to have $P = 1.178$ days. Meanwhile, they were
not able to rule out conclusively or confirm the presence of the 7th
planet. Our analysis of the residual data of the 6-planet model did
reveal several peaks in the posterior distribution with periods around
6.51 and 1 days, but as can be seen from the value of residual
evidence in Tab.~\ref{tab:Z_HD10180}, they were not found to be
sufficiently significant. We therefore rule out the presence of
any additional planets contributing to the RV data.

%%%%%%%%%%%%%%%%%%%%%%%%%%%%%%%%%%%%%%%%%%%%%%%%%%%%%%%%%
\section{Conclusions}\label{sec:conclusions}
%%%%%%%%%%%%%%%%%%%%%%%%%%%%%%%%%%%%%%%%%%%%%%%%%%%%%%%%%

We have presented a new and efficient method to detect extrasolar
planets from RV measurements. Our method is not only able to fit for a
specific number of planets, but can also infer the number of planets
from the data using Bayesian model selection. We have successfully
applied our method to simulated data-sets, as well as to the real
systems HD 37124, 47 Ursae Majoris and HD 10180. Our method can
potentially identify many undiscovered extrasolar planets in existing
RV data-sets. One drawback of our method is that it ignores the
planet-planet interactions, but these interactions are important only
for a very small fraction of planetary systems. Moreover, our basic
methodology can be extended to include such interactions. This will be
undertaken in further work.

Another important avenue of research in extrasolar planet searches is
to perform a coherent analysis using different data-sets, e.g. by
jointly analysing the RV data and light curves for the same
system. This would enable us to place better constraints on the
planetary parameters and also to learn about the physical structure of
the planets. Once again our basic analysis technique can be easily
extended to perform a joint analysis of data sets of different types
We plan to extend our approach by incorporating light curve data in a
forthcoming paper.

%%%%%%%%%%%%%%%%%%%%%%%%%%%%%%%%%%%%%%%%%%%%%%%%%%%%%%%%%
\section*{Acknowledgements}
%%%%%%%%%%%%%%%%%%%%%%%%%%%%%%%%%%%%%%%%%%%%%%%%%%%%%%%%%

We would like to thank the referee, Phil Gregory, for useful comments on the paper and Pedro Carvalho for useful
discussions regarding multiple object detection. This work was carried out largely on the {\sc Cosmos} UK
National Cosmology Supercomputer at DAMTP, Cambridge and the Darwin Supercomputer of the University of Cambridge
High Performance Computing Service ({\tt http://www.hpc.cam.ac.uk/}), provided by Dell Inc. using Strategic
Research Infrastructure Funding from the Higher Education Funding Council for England. FF is supported by a
Research Fellowship from Trinity Hall, Cambridge. STB acknowledges support from the Isaac Newton Studentship.

\bibliographystyle{mn2e}
\bibliography{references}

\appendix

\label{lastpage}

\end{document}